\documentclass[a4paper,11pt]{article}
\usepackage[margin=1in]{geometry}

\usepackage{graphicx}
\usepackage{amsthm}
\newtheorem{theorem}{Theorem}

\usepackage{hyperref}
\usepackage{authblk}

\title{In-Place Initializable Arrays}

\author[1]{Takashi Katoh}
\author[1]{Keisuke Goto}
\affil[1]{Fujitsu Laboratories Ltd., Kawasaki, Japan.\\
	$\mathtt{kato.takashi\_01@fujitsu.com, goto.keisuke@fujitsu.com}$}
\date{}                     
\usepackage{hyperref}
\usepackage{appendix}
\usepackage{enumitem}
\usepackage{cite}
\usepackage[whole]{bxcjkjatype} 
\usepackage{xcolor}
\usepackage[final,mode=multiuser]{fixme}
\usepackage{graphicx}
\usepackage[ruled, vlined, linesnumbered]{algorithm2e}
\usepackage{csvsimple} 
\usepackage{tabularx} 

\fxsetup{theme=color}
\fxuseenvlayout{colorsig}
\fxusetargetlayout{color}
\FXRegisterAuthor{kg}{akg}{KG}
\definecolor{kgnote*}{rgb}{1.0, 0, 0}
\colorlet{kgnotebg*}{red}
\colorlet{kgnotebg}{red}
\fxsetface{env}{}
\fxsetface{inline}{\bfseries}
\fxsetface{margin}{\tiny}

\newcommand{\ii}{i}

\newcommand{\kk}{k}

\newcommand{\Array}[1]{\textbf{#1}}
\newcommand{\T}{\Array{T}}
\newcommand{\F}{\Array{F}}
\newcommand{\A}{\Array{A}}
\newcommand{\B}{B}

\newcommand{\X}{\Array{X}}
\newcommand{\Y}{\Array{Y}}

\newcommand{\floor}[1]{\lfloor #1 \rfloor}
\newcommand{\ceil}[1]{\lceil #1 \rceil}
\newcommand{\mymod}{\mbox{ mod }}

\newcommand{\N}{N}

\newcommand{\one}{1\xspace}

\newcommand{\Z}{\Array{Z}}
\newcommand{\Zd}{\Z}

\newcommand{\V}{\Array{V}}

\newcommand{\None}{\mathit{None}}

\newcommand{\celbit}{\ell}
\newcommand{\numw}{\mathit{b}}
\newcommand{\initv}{\mathit{initv}}
\newcommand{\ip}{{\ii^\prime}}
\newcommand{\flag}{\mathit{flag}}

\newcommand{\func}[1]{{\tt #1}}
\newcommand{\opinit}{\func{init}}
\newcommand{\opread}{\func{read}}
\newcommand{\opwrite}{\func{write}}

\newcommand{\breakChain}{\func{breakChain}}
\newcommand{\extend}{\func{extend}}
\newcommand{\chainOf}{\func{chainedTo}}
\newcommand{\makeChain}{\func{makeChain}}
\newcommand{\initBlock}{\func{initBlock}}

\newcommand{\Ns}{\N^\prime}
\newcommand{\celbits}{\celbit^\prime}
\newcommand{\numpack}{p}

\newcommand{\Ad}{\A}

\newcommand{\logN}{\ceil{\log \N}}
\newcommand{\UCA}{UCA\xspace}
\newcommand{\WCA}{WCA\xspace}

\newcommand{\figwidth}{14cm}

\begin{document}

  \maketitle

\begin{abstract}
An initializable array is an array that supports the read and write operations for any element and the initialization of the entire array.
This paper proposes a simple in-place algorithm to implement an initializable array of length $\N$ containing $\celbit \in O(w)$ bits entries in $\N \celbit +1$ bits on the word RAM model with $w$ bits word size, i.e., the proposed array requires only \one extra bit on top of a normal array of length $\N$ containing $\celbit$ bits entries.
Our algorithm supports the all three operations in constant worst-case time, that is, it runs in-place using at most constant number of words $O(w)$ bits during each operation.
The time and space complexities are optimal since it was already proven that there is no implementation of an initializable array with no extra bit supporting all the operations in constant worst-case time [Hagerup and Kammer, ISAAC 2017].
Our algorithm significantly improves upon the best algorithm presented in the earlier studies [Navarro, CSUR 2014] which uses $\N + o(\N)$ extra bits to support all the operations in constant worst-case time.
\end{abstract}

\section{Introduction}

Arrays are important data structures that support the fundamental read and write operations of any given element in constant worst-case time.
Another fundamental operation known as \textit{initialization}, which writes a given initial value to all the elements of the array, appears frequently in numerous algorithms and programs.
Although initialization is naively implemented by linear time write operations, the naive initialization may cause a bottleneck in the applications which use large arrays and frequently require initialization.
The issue motivates us to study a fundamental data structure \textit{initializable array}.

An initializable array $\Zd[0 \ldots \N-1]$ of length $\N$ containing $\celbit$ bits entries supports the following operations: \textit{read}, \textit{write}, and \textit{initialization}, where $i$ and $v$ are the integers within $0\leq i < N$ and $0 \leq v < 2^\celbit$. 
\begin{itemize}
	\item $\opread(i)$: Return the value stored in the $i$-th element of $\Zd$.
    \item $\opwrite(i, v)$: Set the $i$-th element of $\Zd$ to $v$.
    \item $\opinit(v)$: Set all the elements of $\Zd$ to $v$.
\end{itemize}
$\opread(i)$ and $\opwrite(i, v)$ are also denoted by $\Zd[i]$, $\Zd[i] \leftarrow v$, respectively.

A normal array is obviously an initializable array since it fundamentally supports the $\opread$ and $\opwrite$ operations in constant worst-case time, and $\opinit$ can be implemented by calling $\opwrite$ for all $\N$ positions in $\Theta(\N)$ time.
Note that $\opinit$ does not necessarily have to call $\opwrite$ $\N$ times, and it only has to behave as if it does that.
That is, when $\opread(i)$ is called, it only has to return the initial value $v$ of the last initialization $\opinit(v)$ if $\opwrite(i, v^\prime)$ was not called after the last initialization; otherwise, it returns $v^\prime$ of the last write $\opwrite(i, v^\prime)$.

We assume the word RAM model with $w \in \Omega(\log N)$ bits word size that usual arithmetic and bitwise operations on a word take constant worst-case time, and we also assume that $\celbit \in O(w)$. 
We focus on and evaluate the additional extra space over $\N \celbit$ bits because the $\N \celbit$ bits space is the trivial lower bound.
Moreover, we account only the dynamic values for the space of algorithms, e.g., the space for an initial value or auxiliary arrays.
Conversely, we do not account static values that can be embedded into a program, e.g., the space for the length of the array or certain static parameters of algorithms.

Initializable arrays have been studied since the 1970s. 
A folklore algorithm supporting all the operations in constant worst-case time was first mentioned (but not described) in the study by Aho et al.~\cite[Ex. 2.12]{AhoHU74}.
The complete description was later presented in the studies by Mehlhorn~\cite[Sec. III.8.1]{Mehlhorn84} and Bentley~\cite[Column 1]{Bentley86}.
The most technical point of implementing the efficient initializable arrays is how to memorize information whether each element of $\Z$ was overwritten after the last initialization.
The folklore algorithm memorizes the information by using the chain technique, which represents bi-directional links in two auxiliary arrays.
It requires $2\celbit(\N+1)$ extra bits.
Navarro~\cite{Navarro12, Navarro13}  reduced the space to $\N + o(\N)$ extra bits without increasing the time complexities.
Their algorithm combined the folklore algorithm with a bitmap technique using a bit array $\Array{B}$ of length $\N$ such that $\Array{B}[i]$ is $1$ if and only if the $i$-th element of the array has been written from the last initialization.

The runtime of an algorithm depends on the access frequency to the array, which is the ratio of the number of read and write operations, to the array length.
Fredriksson and Kilpel\"ainen~\cite{FredrikssonK16} measured the runtime performances of several algorithms.
According to their computational experiments, the folklore algorithm and Navarro's algorithm present the highest efficiency when the access frequency is low (below 1\%), while the bitmap solution and the naive solution present the highest efficiency when the access frequency is within 1--10\% and greater than 10\%, respectively.

The construction of ZDD~\cite{Minato93} is a good example to demonstrate the effectiveness of the initializable arrays.
ZDD is a space-efficient data structure which represents any family of sets and is widely used for various practical applications~\cite{Sasao14,Minato17}.
Knuth~\cite{simpath,KnuthTheArt} used the folklore algorithm to implement a large hash table in the fast ZDD construction algorithm \textit{Simpath}.
A hash table is used to represent million of nodes of ZDD and is initialized before each step of the algorithm.
Initializable arrays realize the efficient initialization of such a large hash table.

\paragraph{Our Contributions}

We propose a simple in-place algorithm for initializable arrays, and we show the following theorem.

\begin{theorem}
	\label{thm:one-bit}
	There exists an initializable array $\Zd$ of length $\N$ containing $\celbit \in O(w)$ bits entries which requires \one extra bit and supports the operations, read, write, and initalization in constant worst-case time.
\end{theorem}

The algorithm uses a novel in-place chain technique which is nearly identical to the folklore algorithm but works in-place.
The time and space complexities are optimal since there is no implementation of  an initializable array with no extra bit supporting all the operations in constant worst-case time~\cite{HagerupK17}.
Moreover, the algorithm is extremely simple and the pseudo-code of the core idea is written within 80 lines (see Algorithm~\ref{algo:write}--\ref{algo:tools-others}).

\paragraph{Recent Works}
\begin{table*}[t]
  \centering
  {
  \footnotesize
  \begin{tabular}{|c|c|c|c|c|}
\hline
	Algorithms & Extra bits & $\opinit$ & $\opread$ & $\opwrite$ \\ \hline
	Normal array & 0 & $\Theta(\N)$  & $O(1)$  & $O(1)$  \\ \hline
	Folklore~\cite{AhoHU74,Mehlhorn84,Bentley86} & $2\celbit(\N +1)$ & $O(1)$  & $O(1)$  & $O(1)$  \\ \hline
	Navarro~\cite{Navarro12, Navarro13} & $\N + o(\N)$ & $O(1)$  & $O(1)$  & $O(1)$  \\ \hline \hline
	Hagerup and Kammer~\cite{HagerupK17} & $\ceil{\N / (w / ct)^t}$ & $O(1)$  & $O(t)$  & $O(t)$  \\ \hline
    Loong et al.~\cite{Loong17} & 1 & $O(1)$  & $O(1)$  & amortized/expected $O(1)$ \\ \hline
	This paper & 1 & $O(1)$  & $O(1)$  & $O(1)$  \\ \hline
\end{tabular}
  }
  \caption{
    Comparison of the time and space complexities between the earlier studies, recent ones but independent from us, and ours.
    In~\cite{HagerupK17}, $c>1$ is a constant value and $t$ is a time and space trade off parameter within $1 \leq t \leq \logN$, and the algorithm requires \one extra bit when setting $t=\logN$.
    Only $\opwrite$ by Loong et al. takes amortized time or worst-case expected time, and the other operations take worst-case time.
  }
  \label{table:comparison}
\end{table*}
Hagerup and Kammer~\cite{HagerupK17}, and Loong et al.~\cite{Loong17} have also proposed in-place algorithms for initializable arrays recently and independently from us.
Hagerup and Kammer's algorithm supports the read/write operations in $O(t)$ worst-case time, the initialization in constant worst-case time using extra $\ceil{\N / (w / ct)^t}$ bits, where $c$ is a constant value greater than 1, and $t$ is a time and space trade-off parameter within $1 \leq t \leq \logN$.
The in-place algorithm of \one extra bit space is obtained by setting $t=\logN$, but the read and write operations take $O(\log \N)$ worst-case time.
Loong et al. proposed two algorithms, both of which use \one extra bit and support the read/initialization operations in constant worst-case time, and for write, one of which, takes amortized constant time and the other takes constant worst-case expected time.
Compared to these algorithms, our algorithm is quite simple and runs in optimal time and space.
See also Table~\ref{table:comparison}.

Several space-efficient algorithms have been proposed based on the preprint version of this paper~\footnote{The preprint version of this paper is available at \url{https://arxiv.org/abs/1709.08900}}.
Kammer and Sajenko~\cite{KammerS18} have extended our algorithm to implement \textit{dynamic} initializable arrays which can increase and decrease the array size.
Hagerup~\cite{Hagerup19choice,Hagerup19BFS} used the in-place chain techniques presented in this paper to implement space-efficient choice dictionaries 
which can return an arbitrary element stored after the initialization. 
These studies for highly space-efficient data structures were introduced in~\cite{Hagerup19succ}.

\paragraph{Organizations}
The rest of the paper is organized as follows.
Section~\ref{sec:folklore} introduces the folklore algorithm which our algorithm is based on.
Section~\ref{sec:algo} considers a simple problem setting $\celbit \geq \logN$ and $\celbit \in O(w)$, and proposes in-place algorithm using $2\celbit$ extra bits for this problem setting.
Section~\ref{sec:opt} considers a more general problem setting $\celbit \in O(w)$, and presents the proof of Theorem~\ref{thm:one-bit}.
\section{Folklore Algorithm}
\label{sec:folklore}
The folklore algorithm implements an initializable array $\Zd$ of length $\N$ containing $\celbit$ bits entries for $\celbit \geq \logN$ and $\celbit \in O(w)$, which supports all the operations in constant worst-case time.
The algorithm uses three normal arrays of length $\N$ containing $\celbit$ bits entries, $\V$, $\F$, and $\T$~\footnote{$\V$, $\F$, and $\T$ stand for Value, From, and To, respectively.}, along with two variables of $\celbit$ bits, an initial value $\initv$ and a stack pointer $\numw$, and it thus requires $2\celbit (\N + 1)$ extra bits in total.
$\initv$ stores the initial value, $\T$ is used as a stack, and $\numw$ indicates the stack size of $\T$.
We say that $\F[i]$ and $\T[j]$ are chained when they are linked to each other, namely, $\F[i]=j$, $\T[j]=i$, and $j < \numw$.
$\V[i]$ stores a written value, and we maintain the invariant that $\Z[i]=\V[i]$ if $\F[i]$ is chained, and $\Z[i]=\initv$ otherwise.

The algorithm implements each operation using the invariant as follows:
\begin{itemize}
	\item $\opread(i)$: Return $\V[i]$ if $\F[i]$ is chained, and $\initv$ otherwise.
	\item $\opwrite(i, v)$: Set $\V[i]$ to $v$, and if $\F[i]$ is unchained, create a new chain between $\F[i]$ and $\T[\numw]$ by setting $\T[\numw] \leftarrow i$, $\F[i] \leftarrow \numw$, and $\numw \leftarrow \numw + 1$.
  \item $\opinit(v)$: Break all chains by setting $\numw \leftarrow 0$ and update the initial value $\initv \leftarrow v$.
\end{itemize}
$\opread$ is trivially obtained from the invariant.
$\opwrite$ creates a new chain of $\F$ and $\T$ only when an element is written for the first time, and thus the number of chains is at most $\N$, and the chain will never be broken until $\opinit$ is called.
$\opinit$ breaks all the chains by setting $\numw \leftarrow 0$, and thus it implies that all the elements of $\Z$ are initialized by a new initial value $\initv$.
Each operation takes constant worst-case time.
The folklore algorithm thus maintains the invariant and implements an initializable array $\Z$ using $2 \celbit (\N + 1)$ extra bits, and supports all the operations in constant worst-case time.

\section{In-Place Initializable Arrays for a Simple Setting}
\label{sec:algo}

We propose an algorithm which implements an initializable array $\Zd$ for $\celbit \geq \logN$ and $\celbit \in O(w)$.
The algorithm uses one normal array $\Ad$ of $\N \celbit$ bits and two variables, an initial value $\initv$ of $\celbit$ bits and a stack pointer $\numw$ of $\celbit$ bits; it thus requires $2\celbit$ extra bits.
Section~\ref{sec:opt} then shows that the algorithm can be modified to run in optimal time and space for a more general setting.
In the rest, we only consider the case $\N$ is even since, if $\N$ is odd, we just treat $\Zd[\N-1]=\Ad[\N-1]$.

The underlying concept of our algorithm is nearly identical to that of the folklore algorithm.
Our algorithm also uses $\V$, $\F$, and $\T$, but sparsely embeds them into $\A$.
This idea intuitively seems impossible because all $3\N$ elements of $\V$, $\F$, and $\T$ are required in the worst case in the folklore algorithm, and the space of $\A$ is not sufficient to store all of them.
Hence, we reduce the number of chains to solve this issue.
Firstly, we split $\A$ into $\N/2$ blocks of block size $2$ and create chains between two blocks instead of two elements.
Secondly, we also split $\A$ into two areas $\A[0 \ldots 2\numw-1]$ and $\A[2\numw\ldots \N-1]$, and manage written and unwritten blocks in a different manner.
In the first area, a block is chained if and only if the elements of the block has not been written from the last initialization.
In the second area, a block is chained if and only if the elements of the block has been written from the last initialization.
These two areas are called \textit{unwritten chained area} (\UCA) and \textit{written chained area} (\WCA), respectively.

This idea is derived from the important observation that if the written elements are managed by chains (like the folklore algorithm), a few chains are required at the beginning after the last initialization, but this increases gradually and eventually reaches $\N$.
Conversely, if the unwritten elements are managed by chains, a few chains are required at the ending after the last initialization, but approximately $\N$ chains are required at the beginning.
Our algorithm uses these two different management approaches in two areas of $\A$ by changing the size of the area dynamically.
Here, the threshold of the areas $2\numw$ is set to a position such that the number of chains is the least, namely, the number of unwritten blocks in \UCA and the number of written blocks in \WCA are equaled.

\begin{figure*}[tb]
	\centering
	\includegraphics[width=\figwidth]{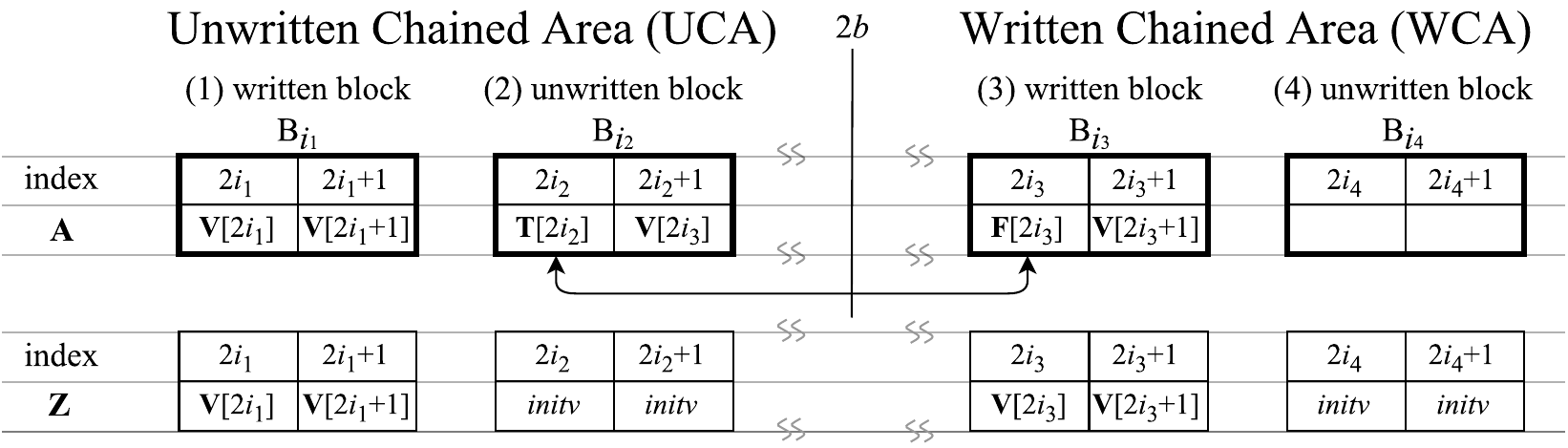}
	\caption{
		Four blocks chained or unchained in the unwritten chained area (\UCA) or the written chained area (\WCA) in $\A$.
		Bold borders indicate blocks.
		Blocks $\B_{i_2}$ and $\B_{i_3}$ are chained since they are in the different areas, $\A[2{i_3}]=\F[2{i_3}]=2{i_2}$ and $\A[2{i_2}]=\T[2{i_2}]=2{i_3}$.
	}
	\label{fig:blocks}
\end{figure*}

The memory layout of $\A$ is shown in Figure~\ref{fig:blocks}.
Let $\B_i=\A[2i \ldots 2i+1]$ be the $i$-th block.
Each $\A[i]$ belongs to the block $\B_{\floor{i/2}}$.
We say that blocks $\B_i$ and $\B_j$ are chained if $\A[2i]= 2j$ and $\A[2j] = 2i$ and neither of the blocks are in the same area.
Note that any element can store any index of $\A$ since $\celbit \ge \logN$.
There are four types of blocks which are classified written or unwritten blocks located in \UCA or \WCA.
For each type of block, our algorithm maintains the following four invariants, where $\V[i]$, $\F[i]$, and $\T[i]$ respectively represent the functional aspects of $\A[i]$ as in the folklore algorithm.

\begin{enumerate}
  \item \label{inv:unchained-written}
  Block $\B_{i_1}$ is a written block in \UCA $\Leftrightarrow$ $\B_{i_1}$ is not chained to any block in \WCA.
  It holds $(\A[2{i_1}], \A[2{i_1}+1])=(\V[2{i_1}], \V[2{i_1}+1])$ and $(\Z[2{i_1}], \Z[2{i_1}+1])=(\V[2{i_1}], \V[2{i_1}+1])$.
  
  \item \label{inv:chained-written}
  Block $\B_{i_2}$ is an unwritten block in \UCA $\Leftrightarrow$ $\B_{i_2}$ is chained to a block $\B_{i_3}$ in \WCA.
  It holds $(\A[2{i_2}], \A[2{i_2}+1])=(\T[2{i_2}], \V[2{i_3}])$ and $(\Z[2{i_2}], \Z[2{i_2}+1])=(\initv, \initv)$. 
  
  \item \label{inv:chained-unwritten}
  Block $\B_{i_3}$ is a written block in \WCA $\Leftrightarrow$ $\B_{i_3}$ is chained to a block $\B_{i_2}$ in \UCA.
  It holds $(\A[2{i_3}], \A[2{i_3}+1])=(\F[2{i_3}], \V[2{i_3}+1])$ and $(\Z[2{i_3}], \Z[2{i_3}+1])=(\V[2{i_3}]=\A[2{i_2}+1], \V[2{i_3}+1])$.

  \item \label{inv:unchained-unwritten}
  Block $\B_{i_4}$ is an unwritten block in \WCA $\Leftrightarrow$ $\B_{i_4}$ is not chained to any block in \UCA.
  It holds $(\Z[2{i_4}], \Z[2{i_4}+1])=(\initv, \initv)$.
\end{enumerate}

The implementation of our algorithm in each operation is described as follows.
$\opread$ is trivially implemented by the invariants.
$\opinit$ is implemented similar to the folklore algorithm by setting $\numw$ and $\initv$ to zero and a given initial value, respectively.
The pseudo-codes of $\opinit$ and $\opread$ are described in Algorithm~\ref{algo:read} in the Appendix.
$\opwrite$ is more complicated than $\opread$ and $\opinit$ since it may create an \textit{unintended chain} by writing a new value and may break the invariants.
For example, if we set $\A[2i_1]$ to $v$ for $\opwrite(2i_1, v)$ in the layout of Figure~\ref{fig:blocks}, it may create unintended chain between $\B_{i_1}$ and $\B_{i_4}$, that is, $\A[2i_1]=2i_4=v$ and $\A[2i_4]=2i_1$.
The following tools~\footnote{
Some of these functions take and return blocks as their arguments and outputs, respectively.
Actual implementations treat such blocks as pointers, so copy and comparison of the constant number of blocks take constant worst-case time.
However, in our pseudo-codes, we represent a block $\B_i$ as just $\B_i$ instead of a pointer $i$ to emphasize that we are indicating a block.} are used to implement $\opwrite$ (their pseudo-codes are shown in Algorithm~\ref{algo:tools-others} that can be found in the Appendix).
\begin{itemize}
	\item $\chainOf(\B_i)$: Return the block chained to $\B_i$ if $\B_i$ is chained, and return a symbol $\None$ otherwise.
	\item $\makeChain(\B_i, \B_j)$: Make a new chain between $\B_i$ in \UCA and $\B_j$ in \WCA.
	\item $\breakChain(\B_i)$: Break the chain of the block $\B_i$ in \UCA if $\B_i$ is chained, and do nothing otherwise.
	\item $\initBlock(\B_i)$: Initialize the block $\B_i$ with $\initv$, namely, write $\initv$ to $\A[2i]$ and $\A[2i+1]$.
	\item $\extend()$: Extend \UCA by one block and return an unwritten block in \UCA that has not yet been chained and is initialized with $(\initv, \initv)$.
\end{itemize}

\newcommand{\numwp}{{s}}
$\chainOf$, $\makeChain$, and $\initBlock$ are directly implemented from their functional aspects.
$\breakChain$ breaks an unintended chain between $\B_i$ and $\B_k$ by setting $\A[2k] \leftarrow 2k$ to keep $B_k$ unchained regardless of the value of $B_i$. %
$\extend$ checks whether the left-most block $\B_{\numwp}$ in \WCA is chained or not, where $\numwp=\numw$.
Let $\B_k$ be the block chained to $\B_{\numwp}$ if it exists.
It then updates $\numw \leftarrow \numw + 1$ ($\numwp$ is unchanged) to extend \UCA, and now $\B_{\numwp}$ has moved form \WCA to \UCA.
(1) If $\B_k$ is none, $\Z[2\numwp]$ and $\Z[2\numwp+1]$ store initial values, so $\B_{\numwp}$ is initialized by calling $\initBlock(\B_{\numwp})$.
This initialization may make an unintended chain between $\B_{\numwp}$ in \UCA and a block $\B_{\initv/2}$ in \WCA, that is, $\initv$ is even, $\initv \ge 2 \numw$, $\A[2\numwp]=\initv$, and $\A[\initv]=2\numwp$.
To fix this unintended chain, we call $\breakChain(\B_{\numwp})$, and then return the unwritten block $\B_{\numwp}$.
(2) If $\B_k$ is not none, $\Z[2\numwp]$ and $\Z[2\numwp+1]$ store some written values, and they are actually stored in $\A[2k+1]$ and $\A[2\numwp+1]$, respectively.
We change the block layouts of $\B_{\numwp}$ and $\B_k$ following the invariants.
We simply write $\Z[2\numwp]=\A[2k+1]$ and $\Z[2\numwp+1]=\A[2\numwp + 1]$ into $\A[2\numwp]$ and $\A[2\numwp+1]$, respectively, and call $\initBlock(\B_k)$.
This change may make unintended chains
for $\B_{\numwp}$ and $\B_k$.
To fix these unintended chains, 
we call $\breakChain(\B_{\numwp})$ and $\breakChain(\B_k)$, and then return the unwritten block $\B_k$.

\begin{algorithm}[tbp]
	\caption{$\opwrite(i, v)$}
	\label{algo:write}
	\SetKw{KwAnd}{and}
	\SetKw{KwOr}{or}
	\SetKwProg{Fn}{Function}{:}{}
		
	\Fn{$\opwrite(i, v)$}{
	
	$\ip \leftarrow \floor{i/2}$  \tcp{$\B_\ip$ is the block that contains $\A[i]$.}
	$\B_k \leftarrow \chainOf(\B_\ip)$ \;
	\uIf{$\ip < \numw$}{
	 \uIf{$\B_k = \None$}{
		\tcp{$\B_\ip$ is a written block in \UCA.}
		 
			$\A[i] \leftarrow v$ \; \label{line:written-unchained-beg} \label{line:write-and-break-beg}
			$\breakChain(\B_\ip)$ \;   \label{line:written-unchained-end}
			 \label{line:write-and-break-end}
		}
		\uElse{
			\tcp{$\B_\ip$ is an unwritten block in \UCA.}
				$\B_j \leftarrow \extend()$ \; \label{line:cond-written-chained-beg}
				
				\uIf{$\B_\ip = \B_j$}{
					
					\tcp{The same situation of just before Line~\ref{line:write-and-break-beg}.}
					We perform the same procedure as in Lines~\ref{line:write-and-break-beg}--\ref{line:write-and-break-end}. \;
				}
				\uElse{
					\label{line:cond-same}
					\tcp{Swap $\B_\ip$ for $\B_j$}
					$(\A[2j], \A[2j+1]) \leftarrow (\A[2\ip], \A[2\ip + 1])$ \;
					$\makeChain(\B_j, \B_k)$ \;
					$\initBlock(\B_\ip)$ \;
					\tcp{The same situation of just before Line~\ref{line:write-and-break-beg}.}
					We perform the same procedure as in Lines~\ref{line:write-and-break-beg}--\ref{line:write-and-break-end}. \;
				}
				\label{line:cond-written-chained-end}
			}
		}
	\uElse{
		\uIf{$\B_k \neq \None$}{
			\tcp{$\B_\ip$ is written block in \WCA.}
			\uIf{$i \bmod 2=0$}{ \label{line:un-chain-beg} \label{line:cond-unwritten-chained-beg}
			     \tcp{Write $v$ to the second element of $\B_k$}
				$\A[2k + 1] \leftarrow v$ \;
			}
			\uElse{
				$\A[i] \leftarrow v$ \tcp{Write $v$ to the second element of~$\B_\ip$}
			} \label{line:un-chain-end} \label{line:cond-unwritten-chained-end}
		}
		\uElse{
			\label{line:cond-just-write}
			\tcp{$\B_\ip$ is an unwritten block in \WCA.}
			$\B_k \leftarrow \extend()$ \; \label{line:cond-unwritten-unchained-beg}
			\uIf{$\B_\ip = \B_k$}{
				\tcp{The same situation of just before Line~\ref{line:write-and-break-beg}.}
				We perform the same procedure as in Lines~\ref{line:write-and-break-beg}--\ref{line:write-and-break-end}. \;
			}
			\uElse{
				$\initBlock(\B_\ip)$ \;
				$\makeChain(\B_k, \B_\ip)$ \;
				\tcp{The same situation of just before Line~\ref{line:un-chain-beg}.}
				We perform the same procedure as in Lines~\ref{line:un-chain-beg}--\ref{line:un-chain-end}. \; \label{line:cond-unwritten-unchained-end}
			}
		}
	}
}
\end{algorithm}

The pseudo-code of $\opwrite$ is described in Algorithm~\ref{algo:write}.
When $\opwrite(i, v)$ is called, there are four major conditions of the block $\B_\ip$ for $\ip = \floor{i/2}$, and we write $v$ while keeping the invariants in each state as follows.

\begin{itemize}
	\item \label{cond:written-unchained}
	$\B_\ip$ is a written block in \UCA (Lines~\ref{line:written-unchained-beg}--\ref{line:written-unchained-end}).\\
	$\Z[i]$ has already been written, and thus we simply rewrite it with a new value $v$.
    We expect $\B_\ip$ is unchained from the invariant, but $\B_\ip$ may be accidentally chained to a block by writing $v$ to $\A[i]$.
    We call $\breakChain(\B_\ip)$ to break such an unintended chain.
	
	\item \label{cond:written-chained}
	$\B_\ip$ is an unwritten block in \UCA and is chained to a block in \WCA(Lines~\ref{line:cond-written-chained-beg}--\ref{line:cond-written-chained-end}).\\
    Since $\B_\ip$ is chained, there is not sufficient space to store $v$.
    To overcome this issue, we extend \UCA, obtain an unwritten block $\B_j$ in \UCA that has not yet been chained, swap $\B_j$ for $\B_\ip$, and write $v$ to $\A[i]$ in the block $\B_\ip$.
	There are two major concerns in the procedure:
	(1) $\B_\ip$ may be equal to $\B_j$ before swapping.
	(2) $\B_\ip$ may be accidentally chained to a block by writing $v$ to $\A[i]$, which is the same situation as in Lines~\ref{line:written-unchained-beg}--\ref{line:written-unchained-end}.
	In case 1, we do not swap $\B_j$ for $\B_\ip$, and simply write $v$ to $\A[i]$.
	In case 2, we break an unintended chain by $\breakChain(\B_\ip)$.
	
	\item \label{cond:unwritten-chained}
	$\B_\ip$ is a written block in \WCA and is chained to a block $\B_k$ in \UCA (Lines~\ref{line:cond-unwritten-chained-beg}--\ref{line:cond-unwritten-chained-end}).\\
	$\Z[i]$ has already been written, and thus we simply write $v$ in the corresponding position $\A[i]$ or $\A[2k+1]$.
	
	\item \label{cond:unwritten-unchained}
	$\B_\ip$ is an unwritten block in \WCA (Lines~\ref{line:cond-unwritten-unchained-beg}--\ref{line:cond-unwritten-unchained-end}).\\
	$\Z[i]$ has been unwritten, and thus we have to make a chain between the block $\B_\ip$ and a block in \UCA.
	We extend \UCA and obtain a new initialized block $\B_k$ in \UCA by calling $\extend$.
	If $\B_\ip=\B_k$, $\B_\ip$ is now located in \UCA and it is the same 	situation as in 
	Lines~\ref{line:write-and-break-beg}--\ref{line:write-and-break-end}, and then we do the same procedure.
	Otherwise, we initialize the block $\B_{\ip}$ and make the chain between $\B_k$ and $\B_{\ip}$.
	It is now the same situation in Lines~\ref{line:un-chain-beg}--\ref{line:un-chain-end}, so we do the same procedure.
\end{itemize}

Roughly speaking, our algorithm extends \UCA (suppressing \WCA) by increasing $\numw$ by one when writing a new value.
This is similar to how a normal array initializes itself by writing a value from left to right.
Our algorithm performs the same operation in a lazy manner, that is, it writes only two values when increasing $\numw$.
In the extreme case, where $2 \numw = \N$, all the elements have already been written, and the contents of $\A$ are completely equal to $\Z$ , that is, $\A[i]=\Z[i]$ for all $0\leq i<\N$.

Therefore, our algorithm maintains the invariants during the operations, and supports all the operations in constant worst-case time using only $2\celbit$ extra bits.

\section{In-Place Initializable Arrays for a General Setting}
\label{sec:opt}

The algorithm in Section~\ref{sec:algo} can be modified so that it requires only \one extra bit and that it runs in a more general problem setting $\celbit \in O(w)$, rather than $\celbit \geq \logN$ and $\celbit \in O(w)$.
Therefore, we have Theorem~\ref{thm:one-bit}.
We firstly describe the former modification in the same problem setting in Section~\ref{sec:algo}, and then describe the latter one.

\begin{figure*}[tb]
	\centering
	\includegraphics[width=\figwidth]{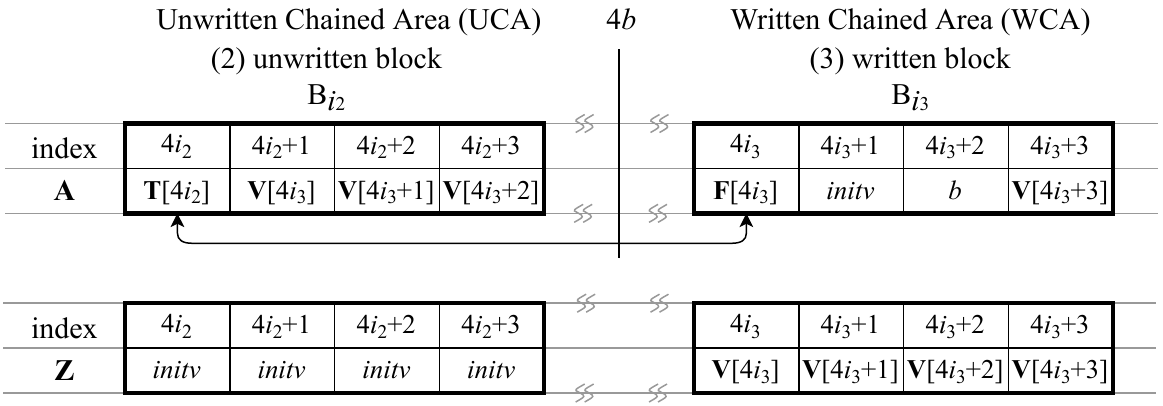}
	\caption{
	The layout of $\A$ and $\Z$ for block size $4$, which is related to Figure~\ref{fig:blocks}.
	The unwritten block $\B_{i_2}$ in \UCA and the written block $\B_{i_3}$ in \WCA are chained.
	The second and third elements of $\B_{i_3}$ can store any $\celbit$ bits values, respectively, and $\B_{i_3}$ in the example stores $\initv$ and $\numw$ in that space.
	Note that we can store any $\celbit$ bit values in the second and third elements regardless of whether the last block is chained or not, and it does not affect the behavior of~$\Z$.
	} 
	\label{fig:opt}
\end{figure*}

We can reduce the space requirement to \one extra bit by changing block size to $4$ and embedding $\initv$ and $\numw$ into the space of the last block in $\A$.
We use only a $1$ bit variable $\flag$, and set $\flag=1$ if and only if the size of \WCA is zero, that is, $\Z=\A$.
If $\flag=1$, we do not need $\initv$ and $\numw$ anymore since $\Z=\A$.
Otherwise, if $\flag=0$, the size of \WCA is not zero, and the last block in $\A$ belongs to \WCA. 
The invariants and the algorithms in Section~\ref{sec:algo} can be easily generalized to the block size greater than $2$.  
See the new layout of the blocks in Figure~\ref{fig:opt}.
From the invariants, the second and third elements of any block in \WCA
do not affect the behavior of $\Z$ regardless of whether the block is chained or not.
Thus, if $\flag=0$, we can store $\initv$ and $\numw$ in that space of the last block.
Our algorithm runs similar to the algorithm of block size $2$ if $\flag=0$, and runs as the normal array if $\flag=1$.
This modification does not worsen the time complexities for all the operations.

\begin{figure*}[tb]
	\centering
	\includegraphics[width=\figwidth]{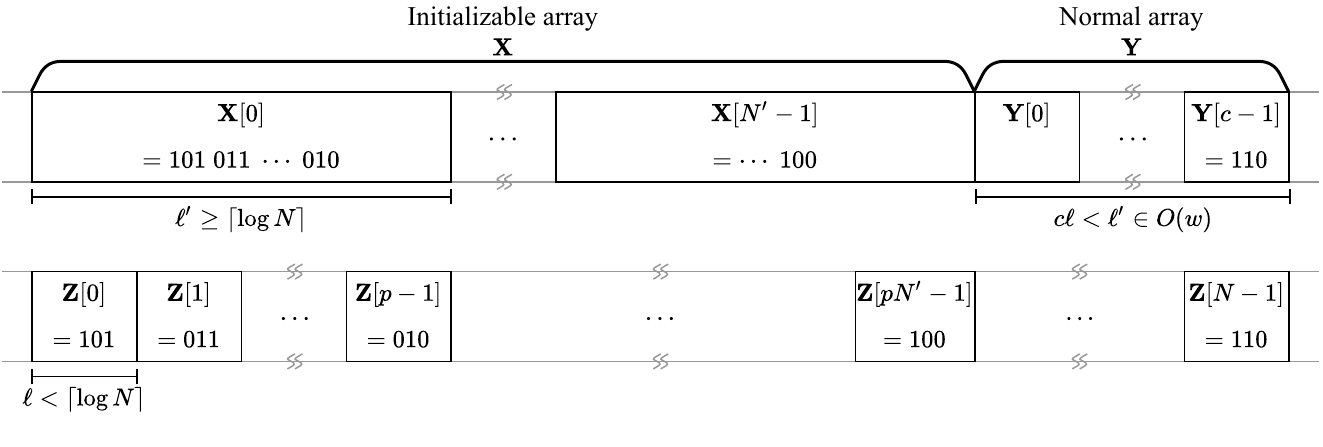}
	\caption{
	An initializabe array $\Z$ of $\celbit < \logN$ bit entries is implemented by an initializable array $\X$ of $\celbits \geq \logN$ bit entries and a normal array $\Y$ of $\celbit$ bit entries.
	The example shows the case $\celbit=3$.
	} 
	\label{fig:layout}
\end{figure*}

\newcommand{\celbitp}{\celbit^\prime}
\newcommand{\Yd}{\Y}
\newcommand{\Xd}{\X}
We can modify the algorithm so that it runs for the more general setting, $\celbit \in O(w)$ rather than $\celbit \geq \logN$ and $\celbit \in O(w)$ without worsening both the time and space complexities.
If $\celbit < \logN$, $\A[i]$ cannot store a pointer to a position in $\A$ and so we cannot use the in-place chain technique as Section~\ref{sec:algo}.
To solve this issue, we simulate an initializable array $\Zd$ by using another initializable array $\X$ containing larger bit size entries and a normal array $\Y$ as shown in Figure~\ref{fig:layout}.
Let $\numpack=\ceil{\frac{\logN}{\celbit}}$, $\Ns = \floor{\frac{\N}{\numpack}}$, $\celbits = \numpack \celbit$, and $c=\N \mymod p=\N - \numpack \Ns$.
$\Xd$ is an initializable array of length $\Ns$ containing $\celbits$ bits entries such that $\Xd[i]=\sum_{j=0}^{p-1}2^{\celbit(p-1-j)} \Zd[ip+j]$ has a bit pattern corresponding to the concatenation of the bit patterns $(\Zd[ip], \ldots, \Zd[(i+1)\numpack - 1])$.
$\Y$ is a normal array of length $c$ containing $\celbit$ bits entries such that $\Y[i]$ is equal to $\Zd[\Ns+i]$ for $0 \leq i < c$.
$\Xd$ can be implemented with only \one extra bit as described earlier since $\logN \leq \celbits < \logN + \celbit$.
We describe only how $\Xd$ simulates $\Zd[0,\ldots,\Ns-1]$ since $\Yd$ can simulate the remaining part of $\Zd$ based on the same concept.
The read and write for $\Zd[i]$ can be performed by reading and writing to $\Xd[\floor{i/\numpack}]$ with the constant number of bit operations.
When $\opinit(\initv)$ is called on $\Zd$, we call $\opinit(\initv^\prime)$ on $\Xd$, where $\initv^\prime$ has a bit pattern corresponding to the concatenation of $p$ consecutive initial values of $\initv$.
Note that $\initv^\prime$ can be computed by multiplying $\initv$ and the pre-computed static bit pattern~\footnote{The pre-computed static bit pattern is embedded within the program.
See also Figure~\ref{fig:initv} in the Appendix.} of length $\celbitp$ whose each $\celbit$-th bit from the left is $1$ and others are $0$.
Therefore, $\Zd$ of $\celbit \in O(w)$ bits entries can be implemented using \one extra bit space, and we have Theorem~\ref{thm:one-bit}.

\section*{Acknowledgements}
We would like to thank Shunsuke Inenaga and Hideo Bannai for many constructive suggestions, and anonymous reviewers for their insightful comments.

  \bibliography{ref}

  \newpage
\section*{Appendix}

\newcommand{\kp}{{\kk^\prime}}

\begin{algorithm}[hb]
	\caption{$\opinit(v)$ and $\opread(i)$}
	\label{algo:read}
	\SetKwProg{Fn}{Function}{:}{}
	\Fn{$\opinit(v)$}{
		$\numw \leftarrow 0$ \;
		$\initv \leftarrow v$ \;
	}
	\Fn{$\opread(i)$}{
	$\ip \leftarrow \floor{i/2}$  \tcp{$\B_\ip$ is the block that contains $\A[i]$.}
	$\B_k \leftarrow \chainOf(\B_\ip)$ \;
	\uIf{$i < 2\numw$}{
		\uIf {$\B_k \neq \None$}{
			\Return $\initv$ \;
		}
		\uElse{\Return $\A[i]$ \;}
		}
	\uElse{
		\uIf{$\B_k \neq \None$}{
			\uIf{$i \bmod 2=0$}{
				\Return $\A[\A[i] + 1]$ \;
				}
			\uElse{\Return $\A[i]$ \;}
		}
		\uElse{\Return $\initv$ \;}
	}
}
\end{algorithm}

\begin{algorithm}[tb]
	\caption{Tools}
	\label{algo:tools-others}
	\SetKwProg{Fn}{Function}{:}{}
	
	\Fn{$\chainOf(\B_i)$}{
		\SetKw{KwAnd}{and}
		\SetKw{KwOr}{or}
		
		$\kp \leftarrow \A[2i] $ \;
		$\kk \leftarrow \floor{\kp/2}$ \tcp{$\B_\kk$ is the block that contains $\A[\kp]$.}
		\uIf {$\kp \bmod 2 = 0$ \KwAnd $(0 \leq i < \numw \leq \kk < \N/2$ \KwOr $0 \leq \kk < \numw \leq i)$ \KwAnd $\A[\kp] = 2i$ }{
			\Return $\B_{\kk}$ \;
		}
		\uElse{
			\Return $\None$ \;
		}
	}
	
	\Fn{$\makeChain(\B_i, \B_j)$}{
		$\A[2i] \leftarrow 2j$ \;
		$\A[2j] \leftarrow 2i$ \;
	}
	
	\Fn{$\breakChain(\B_i)$}{
		$\B_k \leftarrow \chainOf(\B_i)$ \;
		\uIf{$\B_k \neq \None$}{$\A[2k]=2k$ \;}
		}
	
	\Fn{$\initBlock(\B_i)$}{
		$\A[2i] \leftarrow \initv$ \;
		$\A[2i+1] \leftarrow \initv$ \;
	}
	
	\Fn{$\extend()$}{
	    $\numwp \leftarrow \numw$ \;
		$\B_k \leftarrow \chainOf(\B_{\numwp})$ \;
		$\numw \leftarrow \numw + 1$ \;
		
		\uIf{$\B_k = \None$}{
    		$\initBlock(\B_{\numwp})$ \;
    		$\breakChain(\B_{\numwp})$ \;
    		\Return $\B_{\numwp}$ \;
		}
		\uElse{
			$\B_{\numwp} \leftarrow (\A[2k+1], \A[2 \numwp + 1])$\;
			$\breakChain(\B_{\numwp})$ \;
    		$\initBlock(\B_k)$ \;
    		$\breakChain(\B_k)$ \;
		    \Return $\B_k$ \; 
		}
	}
		
\end{algorithm}

\begin{figure*}[htb]
	\centering
	\includegraphics[width=44mm]{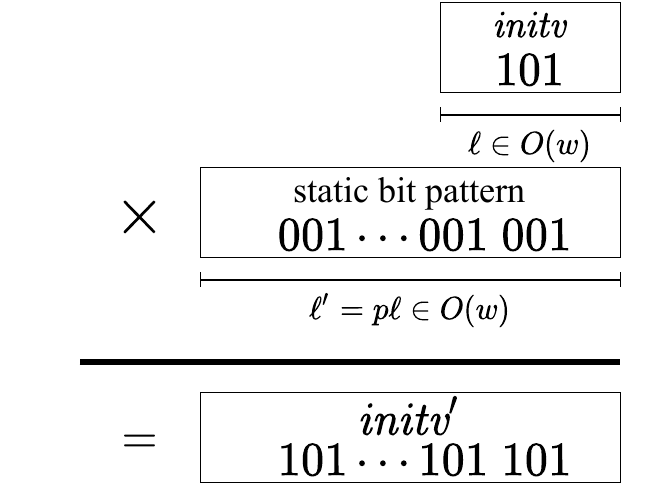}
	\caption{
	The initial value $\initv^\prime$ of $\celbitp \in O(w)$ bits can be computed by multiplying $\initv$ of $\celbit \in O(w)$ bits and pre-computed static bit pattern of $\celbitp$ bits in constant worst-case time.
	The example shows the case $\celbit=3$.
	} 
	\label{fig:initv}
\end{figure*}
\end{document}